\documentclass[preprint,prl,aps,,groupedaddress,showpacs]{revtex4}

\usepackage{graphicx}

\begin{document}

\title{Strain induced half-metal to semiconductor transition in GdN}

\author{Chun-gang Duan, R. F. Sabiryanov, Jianjun Liu,
W. N. Mei}
\affiliation{Department of Physics, University of
Nebraska at Omaha, Omaha, Nebraska 68182-0266}
\author{P. A. Dowben, J. R. Hardy}
\affiliation{Department of Physics and Center for Materials
Research and Analysis, University of Nebraska at Lincoln, Lincoln,
Nebraska 68588}

\date{\today}

\begin{abstract}

We have investigated the electronic structure and magnetic
properties of GdN as a function of unit cell volume. Based on the
first-principles calculations of GdN, we observe that there is a
transformation in conduction properties associated with the volume
increase: first from half-metallic to semi-metallic, then
ultimately to semiconducting. We show that applying stress can
alter the carrier concentration as well as mobility of the holes
and electrons in the majority spin channel. In addition, we found
that the exchange parameters depend strongly on lattice constant,
thus the Curie temperature of this system can be enhanced by
applying stress or doping impurities.
\end{abstract}

\pacs{75.10.-b, 71.27.+a, 71.70.Gm}

\maketitle


Electronic and transport properties of the rare-earth nitrides
have long been a challenge to investigators: the nitrides are
difficult to fabricate into single phase crystals and the
experimental picture of their electronic structures is far from
clear. Although most rare-earth nitrides have shown to be
semi-metallic, there are still uncertainties about GdN. Based on
the direct resistivity measurements, Xiao \emph{et al.} concluded
that GdN is an insulator \cite{XiaoPRL}. Nevertheless, the values
they obtained for the resistivity do not rule out the possibility
of a semi-metallic state. Furthermore, there is no clear-cut
picture emerging from a series of studies by Kaldis and co-authors
in 70's and 80's \cite{Kaldis1,Kaldis2,Kaldis3}. They pointed out
that the specific resistivity decreased with increasing
temperature, which suggests that the GdN system is semiconducting
\cite{Kaldis1}, while their later experiments on better quality
samples showed a very large carrier concentration of 1.9 $\times$
10$^{21}$ cm$^{-3}$, which is more characteristic of a semi-metal
\cite{Kaldis2}. Even though the optical absorption edge for GdN is
about 1 eV \cite{Kaldis3}, which is similar to other rare-earth
nitrides, at issue is the position of the chemical potential
relative to the band edges in both the spin majority and spin
minority band structures.

An appealing property of GdN is that it is ferromagnetic with a
large gap at the Fermi energy in the minority spin states,
according to the electronic structure calculations based on the
local density approximation (LDA) \cite{HY,PLS,LamPRB}. At the
same time GdN is semi-metallic in majority spin states with
electron and hole pockets at the Fermi surface \cite{PLS}. This
latter property has led to some interest in GdN as a possible
candidate for spin-dependent transport devices \cite{spintronics},
exploiting the spin filter, giant magnetoresistance or tunneling
magnetoresistance effects.

An accurate description of the electronic structure of rare-earth
compounds is a very challenging problem because of their unfilled
4\emph{f} shells \cite{Dowbe1}. Calculations based on local spin
density approximation (LSDA) are well known to underestimate the
band gap in semiconductors. Thus LSDA and similar computational
methods may not be able to correctly describe whether a highly
correlated system, like GdN, is semi-metallic or semiconducting at
the equilibrium volume \cite{PLS}. Nonetheless, if we are
interested in the trend of how the electronic and magnetic
properties vary with the change of volume, we can obtain a
reasonable picture for GdN from LSDA with additional Hubbard
correlation terms describing on-site electron-electron repulsion
associated with the 4$f$ narrow bands (LSDA+\emph{U} approach)
\cite{LSDA+U}. Actually, due to the fact that the $f$ states of
this system are exactly half occupied, there is no orbital moment
and the anisotropic and multipole effects are minimal. As a
result, GdN is the ideal material to study the magnetic exchange
interactions in rare-earth nitrides \cite{LiGd}.

In this letter, we show that applying stress can influence
significantly the electronic and magnetic properties of GdN. Using
the first-principles approaches, we demonstrate that the system
exhibits nominal ``half-metallic'' band structure at the
equilibrium lattice constant, and then semi-metallic and/or
semiconducting character develops with increasing lattice
constant. We note that the magnetic properties are also extremely
sensitive to the volume variations, i.e., the exchange
interactions are at first ferromagnetic, then the calculated
magnitudes of exchange parameters reduce substantially with
increasing volume, suggesting that the Curie temperature is
reduced with an increase in lattice constant.

The first-principles band structure approach applied in this work
is the full-potential linear-augmented-plane-wave plus
local-orbital method \cite{WIEN2k}. In the total energy
calculations, the factor $R_{MT}K_{max}$ is chosen to be 8. We
found that using as large as 4000 $k$-points in the Brillouin zone
was necessary to obtain the energy convergence up to 0.1 meV.
Following the previous work \cite{GdJPC}, we used a Hubbard $U$ =
6.7 eV and an exchange $J$ = 0.7 eV for Gd compounds in the
LSDA+\emph{U} scheme \cite{LSDA+U}. The calculated density of
states (DOS) of GdN agrees well with photoemission data of
nitrogen covered Gd(0001) surfaces \cite{Gd0001}, as shown in Fig.
\ref{figDosGdN}. Given that both experiment and theory share the
similar lattice structures, this suggests that our computational
approach correctly represents the electronic structure of GdN.

When we studied the band structure, we found that, for the
theoretical lattice constant $a$ = 4.92 {\AA} [Fig.
\ref{figBandGdN}(a)], our LSDA+\emph{U} calculations show that
there exist a hole pocket at $\Gamma$ point and an electron pocket
at X point. Strong hybridization between Gd 5$d$ and N 2$p$ spin
majority states is clearly shown around the X point, while there
is no such hybridization for spin minority states, indicating GdN
to be half-metallic, with a gap about 0.6 eV in the spin minority
channel. As we increase the lattice constant, this system
gradually develops semi-metallic features in the majority states:
e.g., the $d-p$ hybridization around X point disappears and the
Fermi level down shifts to the top of the hole section at $\Gamma
$ point [Fig. \ref{figBandGdN}(b)]. With further increase of
volume the system eventually becomes semiconducting [Fig.
\ref{figBandGdN}(c)]. The exact volume at which metal-semimetal or
semimetal-semiconductor transitions occurs might not be precisely
determined from LSDA-based calculations. However, the trend is
clear and the predicted transitions could be observed
experimentally.

We note that volume expansion or applied strain can be used to
control the carrier density as well as to some degree the mobility
of the carriers, as can be seen from the band dispersion curves
near the Fermi energy, in Fig. \ref{figBandGdN}. At larger volumes
the electron (hole) pockets become substantially shallower,
meaning that the bottom (top) of the band approaches the Fermi
energy and the area of the Fermi surface decreases. This is quite
clear, for example, when comparing the hole pockets around $\Gamma
$ point in Fig. \ref{figBandGdN}(a) and (b), especially those
Fermi-level crossing points. As a result, the density of states at
the Fermi energy reduces by about 50{\%} when lattice parameter is
increased by only 5{\%}. And the decrease of $dE_{k}/dk$ (tangent
to the band dispersion) at Fermi level can also be seen from Fig.
\ref{figBandGdN}, indicating that the mobility of the holes
decreases with the increase of the cell volume.

Furthermore, as also can be seen from Fig. \ref{figBandGdN}, the
indirect band gap of GdN can be modified by hydrostatic pressure.
Based on the prediction of electron-hole-liquid theory
\cite{Monnier}, a first-order semiconductor to semimetal
transition starts to take place as the indirect band gap decreases
with decreasing lattice constant. Thus it is conceivable that with
proper control of the indirect gap via external pressure, it is
possible to explore the ground state of this correlated
electron-hole liquid, and tune experimentally the semimetal phase
transition.

We have also explored the modifications of the band structure and
electronic properties due to the biaxial strain, which can be
produced by epitaxially growing the film on the substrate with a
different lattice constant. When the lattice parameter of the
substrate is different from that of equilibrium GdN, the lattice
mismatch can produce strain. In such cases, the lattice symmetry
of GdN becomes tetragonal with $c/a$ ratio less than 1 (larger
substrate lattice constant, tensile strain) or larger than 1
(smaller substrate lattice constant, compressive strain). We
performed total energy calculation on the Poisson ratio $\nu$ for
GdN by applying the biaxial strain in \emph{ab} plane and
observing the variation of \emph{c}, then by comparing the biaxial
strain given as
\[
\frac{\Delta c/c_0 }{\Delta a/a_0 }=-\frac{2\nu }{1-\nu },
\]
we deduced a $\nu $ = 0.2, which is small compared to metals but
similar to the known value of TiN. We found that the trend of band
structure change, due to the biaxial strain, remains qualitatively
the same as that caused by volume strain, which is expected
because of the small Poisson ratio. The unit cell volume of GdN
increases due to biaxial tensile stress, hence the modification of
the band structure near the Fermi energy is similar to that
observed during volume expansion. This may imply that the tensile
stress tends to force the system to be less metallic and decreases
the density of states near Fermi surface. When applying the
compressive strain, the system tends to become more metallic.
Based on our calculations, the minority spin shows considerable
reduction in the band gap with bands cross over Fermi energy with
in-plane compressive strain of 3{\%} and then the system is no
longer half-metallic.

We also found that the change in electronic structure with the applied
stress significantly affects the magnetic properties of GdN,
particularly, the exchange interactions. We analyzed the exchange
interactions using Heisenberg Hamiltonian:
\[
H=-\sum\limits_n {J_n \sum\limits_{i>j} {\vec {S}_i \cdot \vec
{S}_j } },
\]
where $J_{n}$ are exchange parameters and $n$ is the index of the
nearest-neighbor shell. In these calculations, we limited our
considerations to the third nearest neighbor interactions, i.e.,
$n$ runs from 1 to 3. The exchange parameters used in the model
Hamiltonian, are obtained from the first-principles band structure
calculations. To do this, we carried out total energy calculations
on four different magnetic ordering configurations of the fcc
structure. One is ferromagnetic (FM) ordering, and the other three
are anti-ferromagnetic (AFM) orderings, as have been described in
Ref. \cite{JSSmart} and here for convenience are called AFM I, II
and III, respectively. According to the model Hamiltonian, the
total exchange energies per magnetic lattice site of the four
magnetic orderings can be explicitly expressed as:
\[
E_{FM} =E_0 +6J_1 +3J_2 +12J_3,
\]
\[
E_{AFM_I } =E_0 -2J_1 +3J_2 -4J_3,
\]
\[
E_{AFM_{II} } =E_0 -3J_2,
\]
\[
E_{AFM_{III} } =E_0 -2J_1 +J_2 +4J_3,
\]
where $E_{0}$ is the reference energy. Based on these equations and the
results obtained from the total energy calculations, those $J$
parameters can be deduced accordingly. Due to the fact that the energy
differences between these ordering states are generally very small,
extreme care is needed in the calculations. Hence, comparisons are only
made between FM and AFM energies calculated on the same structure and
with the same computational parameters, to avoid any error caused by
the different symmetries or shapes.

Our LSDA+\emph{U} calculation gives the correct ground state for
GdN, i.e., FM ordering. The $J_{n}$ values are, nonetheless, quite
small, which is expected from the low transition temperature. In
addition, we found that these exchange parameters of GdN depend
strongly on the lattice constant (Fig. \ref{figJ123Lat}). When the
lattice constant is increasing, the FM $J_{1}$ decreases and
$J_{2}$ even changes sign, namely from FM to AFM, while $J_{3}$
keeps almost the same. Thus the FM transition temperature $T_{c}$,
which is proportional to the sum of neighboring exchange energies
$12J_{1}+6J_{2}+24J_{3}$ according the mean-field theory, is
sensitive to the change of lattice constants.

It is well known that the oscillatory
Rudermann-Kittel-Kasuya-Yosida-type interaction varies sensitively
with the density of charge carriers. Thus, based on the above
observations, we can see that the enhancement of the exchange
interactions between neighboring magnetic sites, when the lattice
constant decreases, is caused by the increase of the number of
free charge carriers as the GdN system becomes more metallic.
Actually, the strong lattice constants dependence of the $J$
parameters of GdN is a manifestation of strong lattice constants
dependence of its electronic structure, which lies between the
metal and insulator phase. Furthermore, the trend of $J_{2}$ with
the increase of the volume implies the strengthening of an AFM
superexchange interaction when the system becomes less metallic.
This can be understood as a competition between the superexchange
and indirect carrier mediated exchange interactions. Usually the
former is roughly proportional to $t^{2}/U$, where \emph{t} is the
band energy or hopping integral, and does not change too strongly
when the lattice parameter increases. Whereas the later, as we
already discussed, would decrease with the increase of the lattice
constant. Hence this reduction causes an overall increase in the
antiferromagnetic coupling between second-nearest-neighbor Gd
sites. Therefore we can see that the magnetic properties of Gd
nitride are strongly related to their electronic properties. Based
on these findings, we expect that GdN, in which a strong AFM and
FM competition exists, could be AFM \cite{Kaldis3} or even more
complicated structure such as spin-glass \cite{LiDXPRB} when
experiencing different stress.

Monte Carlo (MC) simulations, based on the model Heisenberg
Hamiltonian with \emph{ab initio} derived exchange parameters are
used to obtain the Curie temperature. Same method has been applied
successfully in the study of complex permanent magnetic materials
\cite{renatPRL1997}. The lattice studied in our MC simulation is a
$10a \times 10a \times 10a$ fcc cell (4000 spins) with periodic
boundary conditions, where $a$ is the lattice constant. At
theoretical lattice parameter (\emph{a} = 4.92 \AA) we find the
Curie temperature to be about 38 K, agreed reasonably well with
the experimentally observed 58 K. This agreement is quite
impressive, considering the strong correlated nature of this
system. One possible reason for the underestimate of $T_{c}$ may
arise from neglecting the correlated hopping processes in
one-electron picture, which are higher order processes proposed by
Kasuya and Li to explain of ferromagnetic exchange in GdN
\cite{Kasuya}. In addition, we found that the energy levels of
4\emph{f} states are crucial in determining the accurate exchange
parameters.
Our calculation on equilibrium structure of GdN, based on the
present \emph{U} parameter, gives the unoccupied 4\emph{f} states
5 eV above $E_{F}$ and occupied 4 \emph{f} states 6 eV below,
which is in good qualitative agreement with the experimental
situation \cite{Suga}, providing support for the validity of our
studies of the magnetic properties of GdN.

Ordinarily, we would expect the ferromagnetic ground state to
provide a stabilizing role to the spin-dependent electric
conduction. Thus to enhance the Curie temperature of GdN to
practical range is quite important. As we already mentioned above,
one possible way is to apply stress. Moreover, doping impurities
can also be an alternative: it was reported that by adding few
percent of magnetic ions such as Mn, the $T_{c}$ of GaN or ZnO
well exceeds room temperature range \cite{Coey}, hence we believe
this is a promising remedy for concern that the Curie temperature
of GdN is on the low side.


In summary, we have found that there is a large lattice constant
dependence of the electronic and magnetic properties in the spin
dependent band structure of GdN. This material, with half-metallic
gap about 0.6 eV, exhibits a ferromagnetic ground state, rendering
it an attractive candidate for spintronic devices. Hence we appeal
strongly for experimental efforts to study this interesting
compound and validate our claim.

The authors acknowledge helpful discussions with A.G. Petukhov.
This work was supported by the Office of Naval Research, Nebraska
Research Initiative, the Nebraska-EPSCoR-NSF Grant EPS-9720643 and
Department of the Army Grants DAAG 55-98-1-0273 and DAAG
55-99-1-0106. Computation work was completed utilizing the
Research Computing Facility of the University of Nebraska --
Lincoln.



\newpage

%



\begin{figure}
\caption{Comparison between the calculated DOS (majority spin: solid
line, minority spin: dotted line) at theoretical lattice constant
(\emph{a} = 4.92 \AA) and the photoemission spectra for nitrogen on
Gd(0001). The new photoemission features that are not attributable to
the Gd(0001) substrate compare well with the calculated DOS for GdN.
Features with strong N 2\emph{p} weight are indicated. Binding energies
for experiment are shifted to higher binding energies as expected with
a final state spectroscopy of a correlated electron system, but the
shift is roughly uniform for key features of the photoemission spectra
(taken for both \emph{s} and \emph{p} polarized light).}
\label{figDosGdN}
\end{figure}

\begin{figure}
\caption{Band structure of GdN in the vicinity of the Fermi energy
for three volumes: (a) at the calculated equilibrium lattice
parameter \emph{a} = 4.92 \AA, (b) at the lattice parameter
increased by 5\% (\emph{a} = 5.16 \AA), (c) at the lattice
parameter increased by 14\% (\emph{a} = 5.63 \AA). Solid and
dotted lines represent spin majority and spin minority states,
respectively. The change of conducting properties are indicated by
the change of energy difference between the top (bottom) of the
hole (electron) pockets and the Fermi energy. \label{figBandGdN}}
\end{figure}

\begin{figure}
\caption{Exchange parameters of GdN plotted as function of the lattice
constant; notice the strong volume dependence of $J_1$ and $J_2$.
\label{figJ123Lat}}
\end{figure}





\end{document}